\documentclass[fleqn]{revtex4-2}

\usepackage{amsmath,graphicx,float}

\begin{document}

\title{Spin-orbit interaction in the HD$^+$ ion}
\author{Vladimir I. Korobov$^1$ and Jean-Philippe Karr$^{2,3}$}
\affiliation{$^1$Bogoliubov Laboratory of Theoretical Physics, Joint Institute for Nuclear Research, Dubna 141980, Russia}
\affiliation{$^2$Laboratoire Kastler Brossel, Sorbonne Universit\'e, CNRS, ENS-Universit\'e PSL, Coll\`ege de France, 4 place Jussieu, F-75005 Paris, France}
\affiliation{$^3$Universit\'e d'Evry-Val d'Essonne, Universit\'e Paris-Saclay, Boulevard Fran\c cois Mitterrand, F-91000 Evry, France}

\begin{abstract}
We report on progress in calculation of the spin-orbit interaction for the HD$^+$ molecular ion. This interaction is currently the largest source of theoretical uncertainty in determination of the hyperfine structure of rovibrational transition lines. The corrections of order $m\alpha^7\ln(\alpha)$ are derived and numerically calculated. Theoretical hyperfine intervals are compared with experimental data, and the observed discrepancies are discussed.
\end{abstract}
\maketitle

In recent years several experiments \cite{Alighanbari20,Patra20,Kortunov21} succeeded in measuring ro-vibrational transition lines in the HD$^+$ molecular ion with relative precision of a few ppt. On the theoretical side, the "spin-averaged" transition energies were calculated with relative uncertainties of $1.4\times10^{-11}$ for the pure rotational transition and of $7.5\times10^{-12}$ for vibrational transitions \cite{Korobov17,Korobov21}. The main limitation in theoretical predictions comes from the hyperfine stucture of transition lines \cite{FFK21}, although in recent years there has been significant progress in the calculation of the hyperfine structure both in the $m\alpha^6$ order \cite{Hydrogen20,HFS20} and in higher orders \cite{spin-spin20}.

The main purpose of this work is to fill in the gap in the theory by calculating corrections of the order of $m\alpha^7\ln(\alpha)$ to the spin-orbit interaction. This will allow to get theoretical predictions for the favored hyperfine components of a ro-vibrational transition (which keep the spin configuration of the molecular ion unchanged) without any significant loss of precision with respect to the spin-averaged transition energy.

\section{Nonrelativictic QED}

In our derivation of the QED corrections to the hyperfine sublevels of the state we use the Nonrelativictic QED (NRQED) suggested in \cite{Lepage86,Kinoshita96}. The Lagrangian for the NRQED is expressed in terms of nonrelativistic (two-component) Pauli spinor fields $\psi$ for each of the electron, positron, muon, proton, etc. Photons are treated in the same way as in QED. The Lagrangian is constrained by the natural symmetry requirements such as gauge invariance, hermiticity, time reversal symmetry, parity conservation, Galilean invariance and consequently the rotational invariance. The Coulomb gauge is used for the NRQED calculations.

Following \cite{Paz15} we use operators: $D_t=\partial_t\!+\!ieA_0$, $\mathbf{D}=\boldsymbol{\nabla}\!-\!ie\mathbf{A}$, $\mathbf{B}$, $\mathbf{E}$, $\boldsymbol{\sigma}$, as building blocks of the Lagrangian and expand it into a series of inverse powers of the electron mass $m$:
\begin{equation}
\mathcal{L} = \sum_{n=0} \psi_f^*\frac{O_n}{m_e^n}\psi_e.
\end{equation}
Spatial parity and time reversal symmetries of the operators are given in Table \ref{t:symmetries}.

\begin{table}[t]
\begin{center}
\begin{tabular}{c@{\hspace{5mm}}c@{\hspace{5.5mm}}c@{\hspace{5.5mm}}c@{\hspace{5.5mm}}c}
\hline\hline
    & $\boldsymbol{\pi}$ & \textbf{E} & \textbf{B} & $\boldsymbol{\sigma}$ \\
\hline
$P$ & $-$ & $-$ &  +  &  +  \\
$T$ & $-$ &  +  & $-$ & $-$ \\
\hline
\vrule width0pt height 10pt
Dimension    & $M^1$ & $M^2$ & $M^2$ & $M^0$ \\
\hline\hline
\end{tabular}\\[2mm]
\caption{Spatial parity and time reversal symmetries, and mass dimension of operators.}\label{t:symmetries}
\end{center}
\end{table}

Using symmetries imposed on the Lagrangian, one can show that the form of $\mathcal{L}$ \textbf{is unique}, and the coefficients: $c_F$, $c_D$, etc. can be unambiguously obtained from a comparison with the scattering amplitude in QED after choosing the regularization in NRQED. The only arbitrariness is associated with the choice of a basis for homogeneous polynomials of $\mathbf{p}$, $\mathbf{p'}$ (the electron impulse before and after scattering) to express the interactions in momentum space. Namely, for terms of degree 2, the interactions may be separated into $p^2\!+\!p'^2$, $\mathbf{p}\mathbf{p}'$ or $(\mathbf{p}\!+\!\mathbf{p}')^2$, $(\mathbf{p}\!-\!\mathbf{p}')^2$.

With the help of these rules we arrive at the NRQED Hamiltonian expanded up to terms of $1/m^4$ order:
\begin{equation}\label{NRQED_H}
\begin{array}{@{}l}\displaystyle
H_I = eA_0
   -c_F\frac{e}{2m}\boldsymbol{\sigma}\mathbf{B}
   -c_D\frac{e}{8m^2}\bigl[\boldsymbol{\nabla}\mathbf{E}\bigr]
   +c_S\frac{e}{8m^2}\,
                   \boldsymbol{\sigma}\!\cdot\!
                   \Bigl(
                      \boldsymbol{\pi}\!\times\!\mathbf{E}
                      -\mathbf{E}\!\times\!\boldsymbol{\pi}
                   \Bigr)
\\[3mm]\hspace{12mm}\displaystyle
      +c_W\frac{e}{8m^3}\Bigl\{\boldsymbol{\pi}^2,\boldsymbol{\sigma}\mathbf{B}\Bigr\}
      -c_{q^2}\frac{e}{8m^3}\,
         \boldsymbol{\sigma}\!\cdot\![\Delta\mathbf{B}]
      +c_{p'p}\frac{e}{8m^3}
         \Bigl\{\boldsymbol{\pi}\!\cdot\!\mathbf{B}\>\boldsymbol{\sigma}\!\cdot\!\boldsymbol{\pi}\Bigr\}
\\[3mm]\hspace{20mm}\displaystyle
      +c_M\frac{e}{8m^3}
         \Bigl(
            \boldsymbol{\pi}\!\cdot\![\boldsymbol{\nabla}\!\times\!\mathbf{B}]
            +[\boldsymbol{\nabla}\!\times\!\mathbf{B}]\!\cdot\!\boldsymbol{\pi}
         \Bigr)
     +c_A\frac{e^2}{8m^3}\left(\mathbf{E}^2-\mathbf{B}^2\right)
\\[3mm]\hspace{12mm}\displaystyle
      +c_{X_1}\frac{e}{128m^4}\left[\boldsymbol{\pi}^2,(\mathbf{D}\mathbf{E}+\mathbf{E}\mathbf{D})\right]
      +c_{X_2}\frac{e}{64m^4}\,\Bigl\{\boldsymbol{\pi}^2,\left[\boldsymbol{\nabla}\mathbf{E}\right]\Bigr\}
      -c_{X_3}\frac{e}{8m^4}\,\Bigl[\Delta\left[\boldsymbol{\nabla}\mathbf{E}\right]\Bigr]
\\[3mm]\hspace{20mm}\displaystyle
      -c_{Y_1}\frac{e}{64m^4}\Bigl\{
         \boldsymbol{\pi}^2,\boldsymbol{\sigma}\!\cdot\!
         \Bigl(\boldsymbol{\pi}\!\times\!\mathbf{E}\!-\!\mathbf{E}\!\times\!\boldsymbol{\pi}\Bigr)
      \Bigr\}
      +c_{Y_2}\frac{ie}{4m^4}\>\epsilon_{ijk}\sigma^i \pi^j[\boldsymbol{\nabla}\mathbf{E}]\pi^k\>,
\end{array}
\end{equation}
where $\boldsymbol{\pi}=-i\mathbf{D}=\mathbf{p}-e\mathbf{A}$, $\mathbf{E}=-\partial_t\mathbf{A}\!-\!\boldsymbol{\nabla}A_0$, and $\mathbf{B} = \boldsymbol{\nabla}\!\times\!\mathbf{A}$. Covariant derivatives in square brackets act only on the fields within the brackets. Similar results but in the Lagrangian form were obtained previously in \cite{Manohar97} for terms up to $1/m^3$ order and in \cite{Paz13} including terms of order $1/m^4$.

The coefficients $c_i$ are defined as follows:
\begin{equation}\label{c_i}
\begin{array}{@{}l@{\hspace{5mm}}l@{\hspace{5mm}}l}
c_F = 1+a_e,
&
c_S = 1+2a_e,
&
c_D = 1\!+\!2a_e\!+\!\frac{\alpha}{\pi}\frac{8}{3}\left[\ln\left(\frac{m}{2\Lambda}\right)\!+\!\frac{5}{6}\!-\!\frac{3}{8}\right],
\\[1mm]
c_W = 1,
&
c_{q^2} = \frac{a_e}{2}\!+\!\frac{\alpha}{\pi}\frac{4}{3}\left[\ln\left(\frac{m}{2\Lambda}\right)\!+\!\frac{5}{6}\!-\!\frac{1}{8}\right],
&
c_{p'p} = a_e,
\\[1mm]
c_M = \frac{a_e}{2}\!+\!\frac{\alpha}{\pi}\frac{4}{3}\left[\ln{\frac{m}{2\Lambda}}\!+\!\frac{5}{6}\!-\!\frac{3}{8}\right],
&
c_A = 1\!+\!a_e,
\\[1mm]
c_{X_1} = 5\!+\!4a_e,
&
c_{X_2} = 3\!+\!4a_e,
&
c_{X_3} = \frac{\alpha}{\pi}\left[\frac{11}{15}\ln\left(\frac{m}{2\Lambda}\right)\!+\!\frac{5}{6}\!-\!\frac{13}{40}\right],
\\[1mm]
c_{Y_1} = 3\!+\!4a_e,
&
c_{Y_2} = -\frac{\alpha}{\pi}\frac{1}{3}\left[\ln\left(\frac{m}{2\Lambda}\right)\!+\!\frac{5}{6}\!+\!\frac{1}{8}\right].
\end{array}
\end{equation}
where $a_e$ is the anomalous magnetic moment of an electron, $\Lambda\approx m\alpha^2$ is a cutoff parameter, which determines the upper limit of the photon momenta in the NRQED integrations.

When we are interested in calculating contributions only of order $m\alpha^7\ln(\alpha)$, we may replace $\Lambda$ by $m\alpha^2$ in the coefficients and ignore all higher order contributions, as in the following example:
\[
c_{q^2} \Rightarrow \frac{\alpha}{\pi}\frac{4}{3}\ln\left(\alpha^{-2}\right), \quad \hbox{etc.}
\]

\section{Advanced hyperfine structure theory of HD$^+$}

The effective spin Hamiltonian for the hyperfine splitting of a ro-vibrational state in the molecular ion HD$^+$ may be written in a form \cite{HDplus_HFS06}:
\begin{equation}\label{HFS_eff_h}
\begin{array}{@{}l}
\displaystyle
H_{\rm HFS} =
    E_1(\mathbf{L}\cdot\mathbf{s}_e)
   +E_2(\mathbf{L}\cdot\mathbf{I}_p)
   +E_3(\mathbf{L}\cdot\mathbf{I}_d)
   +E_4(\mathbf{I}_p\cdot\mathbf{s}_e)
   +E_5(\mathbf{I}_d\cdot\mathbf{s}_e)
\\[1mm]\displaystyle\hspace{15mm}
   +E_6\Bigl\{2\mathbf{L}^2(\mathbf{I}_p\cdot\mathbf{s}_e)
          \!-\!3[(\mathbf{L}\cdot\mathbf{I}_p)(\mathbf{L}\cdot\mathbf{s}_e)
            \!+\!(\mathbf{L}\cdot\mathbf{s}_e)(\mathbf{L}\cdot\mathbf{I}_p)]
       \Bigr\}
\\[1mm]\displaystyle\hspace{15mm}
   +E_7\Bigl\{2\mathbf{L}^2(\mathbf{I}_d\cdot\mathbf{s}_e)
          \!-\!3[(\mathbf{L}\cdot\mathbf{I}_d)(\mathbf{L}\cdot\mathbf{s}_e)
            \!+\!(\mathbf{L}\cdot\mathbf{s}_e)(\mathbf{L}\cdot\mathbf{I}_d)]
       \Bigr\}
\\[1mm]\displaystyle\hspace{15mm}
   +E_8\Bigl\{2\mathbf{L}^2(\mathbf{I}_p\cdot\mathbf{I}_d)
          \!-\!3[(\mathbf{L}\cdot\mathbf{I}_p)(\mathbf{L}\cdot\mathbf{I}_d)
            \!+\!(\mathbf{L}\cdot\mathbf{I}_d)(\mathbf{L}\cdot\mathbf{I}_p)]
       \Bigr\}
\\[-1mm]\displaystyle\hspace{15mm}
   +E_9\Bigl[\mathbf{L}^2\mathbf{I}_d^2
          -\frac{3}{2}(\mathbf{L}\cdot\mathbf{I}_d)
          -3(\mathbf{L}\cdot\mathbf{I}_d)^2
       \Bigr].
\end{array}
\end{equation}
Couplings of the nuclear magnetic moments with the orbital magnetic field (coefficients $E_2$, $E_3$, and $E_8$), as well as the deuteron quadrupole moment coupling with the orbital part ($E_9$) are very small (see Table \ref{T:E_i}). They can be treated perturbatively and their values calculated within the Breit-Pauli approximation as in \cite{HDplus_HFS06} are sufficiently accurate for the present level of theoretical precision. Magnitude of other coefficients (in MHz) is also shown in the Table.

We use the coupling scheme for angular momentum operators: $\mathbf{F} = \mathbf{s}_e+\mathbf{I}_p$, $\mathbf{S} = \mathbf{F}+\mathbf{I}_d$, $\mathbf{J} = \mathbf{F}+\mathbf{L}$, which reflects (see Fig.1) the fact that the hyperfine structure is predominantly defined by the spin configuration of the system and interaction of the total spin $\mathbf{S}$ with the total orbital angular momentum is the smallest coupling in the hyperfine splitting of a ro-vibrational level.

Coefficients $E_4$ and $E_5$ had been calculated in \cite{spin-spin20} with a relative uncertainty of about $10^{-6}$. Coefficients $E_6$ and $E_7$ were obtained with account of the terms of $m\alpha^6$ order in \cite{HFS20}. Here we focus on the coefficient $E_1$, which requires contributions of the order of $m\alpha^7\ln(\alpha)$ to be taken into account.

\begin{figure}
\begin{center}
\includegraphics[width=.4\textwidth]{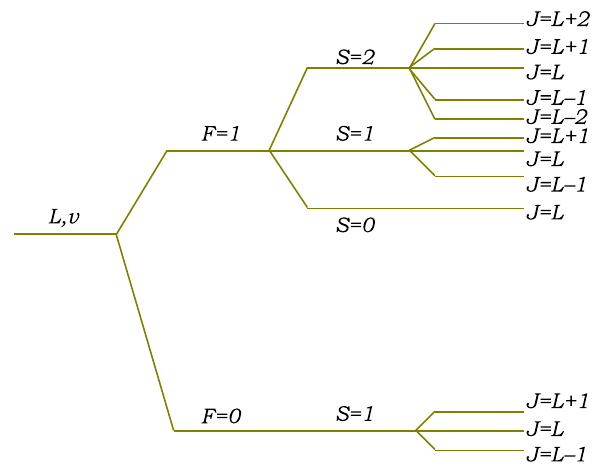}\\
\end{center}
\caption{Schematic diagram of the hyperfine spliting of a rovibrational state (L,v) of HD$^+$.}
\end{figure}

\begin{table}[t]
\begin{center}
\begin{tabular}{@{\hspace{1mm}}c@{\hspace{4mm}}c
@{\hspace{4mm}}c@{\hspace{5.5mm}}c@{\hspace{6mm}}c@{\hspace{6mm}}c@{\hspace{6mm}}c@{\hspace{4mm}}c@{\hspace{4mm}}c}
\hline
\vrule width0pt height9.5pt
 $E_1$ &  $E_2$ & $E_3$ & $E_4$ & $E_5$ & $E_6$ & $E_7$ & $E_8$ & $E_9$ \\
\hline
\vrule width0pt height10pt
31.9846 & $-$3.134[$-$02] & $-$4.809[$-$03] & 924.569 & 142.161 & 8.6111 & 1.3218 & $-$3.057[$-$03] & 5.666[$-$03] \\
\hline
\end{tabular}
\end{center}
\caption{Coefficients $E_i$ of the effective spin Hamiltonian (in MHz) for the $(L\!=\!1,\!v\!=\!0)$ state (from \cite{HDplus_HFS06}), $a[b]=a\times10^{b}$.}\label{T:E_i}
\end{table}

\subsection{Spin-orbit interaction}

The leading-order $m\alpha^4$ relativistic corrections for the spin-orbit interaction is expressed \cite{HFS20}:
\begin{equation}
H_{so} =
   -c_F\left(
         \frac{Z_1}{mM_1}\,\frac{[\mathbf{r}_1\!\times\!\mathbf{P}_1]}{r_1^3}+\frac{Z_2}{mM_2}\,\frac{[\mathbf{r}_2\!\times\!\mathbf{P}_2]}{r_2^3}
      \right)\mathbf{s}_e
   +c_S\left(
         \frac{Z_1}{2m^2}\,\frac{[\mathbf{r}_1\!\times\!\mathbf{p}_e]}{r_1^3}+\frac{Z_2}{2m^2}\,\frac{[\mathbf{r}_2\!\times\!\mathbf{p}_e]}{r_2^3}
      \right)\mathbf{s}_e,
\end{equation}
where coefficients $c_F$ and $c_S$ are defined as in Eq.~(\ref{c_i}), $\mathbf{p}_e$ and $\mathbf{P}_a$ are impulses of the electron and nuclei, correspondingly.

The effective Hamiltonian at $m\alpha^6$ and $m\alpha^7\ln(\alpha)$ orders is derived from the NRQED Hamiltonian (\ref{NRQED_H}) in the same way as in \cite{HFS20}, thus the spin-orbit interaction is expessed by a sum of the following operators:
\begin{equation}
H_{so}^{(6)} = c_{W}\mathcal{U}_{W} + c_{q^2}\mathcal{U}_{q^2}
   + c_{Y_1}\mathcal{U}_{Y_1} + c_{Y_2}\mathcal{U}_{Y_2}
   + c_S\mathcal{U}_{\rm CM} + \mathcal{U}_{\mathrm{MM}_N},
\end{equation}
where the first four operators are obtained from the tree-level contributions:
\begin{subequations}
\begin{equation}
\begin{array}{@{}l}\displaystyle
\mathcal{U}_{W} =
   \frac{Z_1}{4m^3M_1}
   \left\{p_e^2,\frac{1}{r_1^3}\bigl[\mathbf{r}_1\!\times\!\mathbf{P}_1\bigr]\right\}\mathbf{s}_e
   +\frac{Z_2}{4m^3M_2}
   \left\{p_e^2,\frac{1}{r_2^3}\bigl[\mathbf{r}_2\!\times\!\mathbf{P}_2\bigr]\right\}\mathbf{s}_e\,,
\\[1mm]\displaystyle 
\mathcal{U}_{q^2} =
   \frac{iZ_1}{8m^3M_1}\Bigl\{\left[\mathbf{p}_e\!\times\!\left(4\pi\delta(\mathbf{r}_1)\right)\mathbf{P}_1\right]\Bigr\}\mathbf{s}_e
   +\frac{iZ_2}{8m^3M_2}\Bigl\{\left[\mathbf{p}_e\!\times\!\left(4\pi\delta(\mathbf{r}_2)\right)\mathbf{P}_2\right]\Bigr\}\mathbf{s}_e\,,
\\[1mm]\displaystyle
\mathcal{U}_{Y_1} =
   -\frac{Z_1}{16m^4}
         \left\{p_e^2,\frac{1}{r_1^3}[\mathbf{r}_1\!\times\!\mathbf{p}_e]\right\}\mathbf{s}_e
   -\frac{Z_2}{16m^4}
         \left\{p_e^2,\frac{1}{r_2^3}[\mathbf{r}_2\!\times\!\mathbf{p}_e]\right\}\mathbf{s}_e\,,
\\[1mm]\displaystyle
\mathcal{U}_{Y_2} = \frac{iZ_1}{2m^4}\left[\mathbf{p}_e\!\times\!(4\pi\delta(\mathbf{r}_1))\mathbf{p}_e\right]\mathbf{s}_e
   +\frac{iZ_2}{2m^4}\left[\mathbf{p}_e\!\times\!(4\pi\delta(\mathbf{r}_2))\mathbf{p}_e\right]\mathbf{s}_e\,,
\end{array}
\end{equation}
while the last two are from the seagull-type diagrams:
\begin{equation}
\begin{array}{@{}l}\displaystyle
\mathcal{U}_{\rm CM} =
   \frac{Z_1^2}{4m^2M_1}\>
     \frac{1}{r_1^4}\left[\mathbf{r}_1\!\times\!\mathbf{P}_1\right]\mathbf{s}_e
   +\frac{Z_2^2}{4m^2M_2}\>
     \frac{1}{r_2^4}\left[\mathbf{r}_2\!\times\!\mathbf{P}_2\right]\mathbf{s}_e
\\[1mm]\displaystyle\hspace{10mm}
   +\frac{Z_1Z_2}{4m^2M_1}\>
     \frac{1}{r_1r_2^3}\left[\mathbf{r}_2\!\times\!\mathbf{P}_1\right]\mathbf{s}_e
   +\frac{Z_1Z_2}{4m^2M_2}\>
     \frac{1}{r_1^3r_2}\left[\mathbf{r}_1\!\times\!\mathbf{P}_2\right]\mathbf{s}_e
\\[1mm]\displaystyle\hspace{10mm}
   -\frac{Z_1Z_2}{4m^2M_1}\>
     \frac{1}{r_1^3r_2^3}\left[\mathbf{r}_1\!\times\!\mathbf{r}_2\right](\mathbf{r}_1\mathbf{P}_1)\mathbf{s}_e
   +\frac{Z_1Z_2}{4m^2M_2}\>
     \frac{1}{r_1^3r_2^3}\left[\mathbf{r}_1\!\times\!\mathbf{r}_2\right](\mathbf{r}_2\mathbf{P}_2)\mathbf{s}_e\,,
\\[3mm]\displaystyle
\mathcal{U}_{{\rm MM}_N} =
   -\frac{Z_1^2}{2m^2M_1}\>
      \frac{1}{r_1^4}\left[\mathbf{r}_1\!\times\!\mathbf{p}_e\right]\mathbf{s}_e
   -\frac{Z_2^2}{2m^2M_2}\>
      \frac{1}{r_2^4}\left[\mathbf{r}_2\!\times\!\mathbf{p}_e\right]\mathbf{s}_e\,.
\end{array}
\end{equation}
\end{subequations}

The second-order perturbation contributions are expressed as follows:
\begin{equation}
\begin{array}{@{}l}\displaystyle
\Delta E_{so}^{2^{nd}-order} = \Delta E_{so}^{(6)} + \Delta E_{so\hbox{-}ret} + \Delta E_{so\hbox{-}so}^{(1)} + \Delta E_{so}^{(7)},
\\[3mm]\displaystyle
\Delta E_{so}^{(6)} =
   2\left\langle
      H_{so} Q (E_0-H_0)^{-1} Q H_B^{(4)}
   \right\rangle,
\\[2mm]\displaystyle
\Delta E_{so\hbox{-}ret} =
   2\left\langle
      H_{so} Q (E_0-H_0)^{-1} Q H_{ret}
   \right\rangle,
\\[2mm]\displaystyle
\Delta E_{so\hbox{-}so}^{(1)} =
   \left\langle
      H_{so} Q (E_0-H_0)^{-1} Q H_{so}
   \right\rangle^{(1)},
\\[2mm]\displaystyle
\Delta E_{so}^{(7)} =
   2\left\langle
      H_{so} Q (E_0-H_0)^{-1} Q H_B^{(5)}
   \right\rangle.
\end{array}
\end{equation}
Here we include into the Breit-Pauli Hamiltonian radiative corrections (as contribution to the Darwin coefficient, $c_D$, of Eq.~(\ref{NRQED_H})):
\[
\begin{array}{@{}l}\displaystyle
H_B = -\frac{p^4}{8m^3}+c_D\frac{\pi\alpha}{2m^2}\left[Z_1\delta(\mathbf{r}_1)\!+\!Z_2\delta(\mathbf{r}_2)\right] =
   \left\{-\frac{p^4}{8m^3}+\frac{\pi\alpha}{2m^2}\left[Z_1\delta(\mathbf{r}_1)\!+\!Z_2\delta(\mathbf{r}_2)\right]\right\}
\\[2mm]\displaystyle\hspace{28mm}
   +\left(\frac{\alpha}{\pi}\frac{8}{3}\ln{\alpha^{-2}}\right)\frac{\pi\alpha}{2m^2}\left[Z_1\delta(\mathbf{r}_1)\!+\!Z_2\delta(\mathbf{r}_2)\right]
   = H_B^{(4)}+H_B^{(5)}.
\end{array}
\]
It is worth noting that both first and second-order terms are finite and do not require regularization.

\section{Results}

\subsection{Pure rotational transition}

In case of the pure rotational transition \cite{Alighanbari20}: $(L\!=\!0,v\!=\!0)\to(L'\!=\!1,v'\!=\!0)$, the hyperfine splitting is essentially larger relative to the transition line frequency magnitude than in the case of vibrational transitions. Using the six measured transition lines one can extract the experimental value of the $E_1$ coefficient. To do this, we fixed the coefficients of the effective HFS Hamiltonian (\ref{HFS_eff_h}), taking the best theoretical values for $E_i$, and then fit either two parameters: $f_{\rm spin-avg}$ and $E_1$, the spin-averaged transition frequency and the spin-orbit coupling coefficient, or four parameters: $f_{\rm spin-avg}$, $E_1$, $E_6$, and $E_7$. Results obtained by both methods agree well with each other and provide the following numbers:
\[
f_{\rm spin-avg}^{(\rm exp)} = 1\,314\,925\,752.905(10)_{\rm fit}(17)_{\rm exp} \hbox{ kHz},
\]
and
\[
E_1^{\rm exp} = 31984.9(1) \hbox{ kHz}.
\]

Results of the numerical calculations for the spin-orbit coupling coefficient are presented in Table \ref{t:E1}. From comparison with experimental fit one may see that there is some disagreement between theory and experiment of about $5\sigma$. Thus we see that there is room for further careful study of this transition both in theory and experiment.

\begin{table}
\begin{center}
\begin{tabular}{l@{\hspace{12mm}}r}
\hline\hline
  & (1,0)~~~~  \\
\hline
$E_1^{(4)}$ & 31984.645 \\
$E_1^{(6)}$ &     1.119 \\
$E_1^{(7)}$ &  $-$0.347 \\
\hline
$E_1^{\rm th}$  & 31985.4(1)\hspace*{-1mm} \\
\hline
$E_1^{\rm exp}$ & 31984.9(1)\hspace*{-1mm} \\
\hline\hline
\end{tabular}
\end{center}
\caption{Contributions to the $E_1$ coefficient for the $(L\!=\!1,v\!=\!0)$ state (in kHz).}\label{t:E1}
\end{table}

\subsection{Two-photon vibrational transition}

In case of the two-photon vibrational transition: $(L\!=\!3,v\!=\!0)\to(L'\!=\!3,v'\!=\!9)$, the two following lines were measured \cite{Patra20}: $f_0^{\rm HF}$: $(F\!=\!0,S\!=\!1,J\!=\!4)\to(F'\!=\!0,S'\!=\!1,J'\!=\!4)$, and $f_1^{\rm HF}$: $(F\!=\!1,S\!=\!2,J\!=\!5)\to(F'\!=\!1,S'\!=\!2,J'\!=\!5)$. The hyperfine splitting relative to the transition frequency is smaller, so that a precision of five significant digits in the coefficients of the effective HFS Hamiltonian (\ref{HFS_eff_h}) already allows to get a smaller absolute theoretical uncertainty than that of the spin-averaged transition frequency.

Table \ref{E1_Amst} summarizes the contributions to the $E_1$ coefficient both for the initial $(3,0)$ and final $(3,9)$ states. Our final prediction for the $f_{10}^{\rm HF}=f_0^{\rm HF}\!-\!f_1^{\rm HF}$ spitting (see Table \ref{Ams_comparison}) confirms our previous conclusion \cite{HFS20} concerning the disagreement between theory and experiment with a deviation of about $9\sigma$.

\begin{table}
\begin{center}
\begin{tabular}{l@{\hspace{7mm}}r@{\hspace{7mm}}r}
\hline\hline
  & (3,0)~~~~ & (3.9)~~~~ \\
\hline
$E_1^{(4)}$ & 31627.352 & 18270.577 \\
$E_1^{(6)}$ &     1.093 &     0.488 \\
$E_1^{(7)}$ &  $-$0.341 &  $-$0.184 \\
\hline
$E_1$ & 31628.1(1)\hspace*{-1mm} & 18270.9(1)\hspace*{-1mm} \\
\hline\hline
\end{tabular}
\end{center}
\caption{Contributions to the $E_1$ coefficient for the $(L\!=\!3,v\!=\!0)$ and $(3,9)$ states (in kHz).}\label{E1_Amst}
\end{table}

\begin{table}
\begin{center}
\begin{tabular}{l@{\hspace{7mm}}r}
\hline\hline
$f_{10,theo}^{\rm HF}$ & 178\,245.9(0.3) \\
$f_{10,exp}^{\rm HF}$  & 178\,254.4(0.9) \\
\hline\hline
\end{tabular}
\end{center}
\caption{Comparison of the HFS interval $f_{10}$ with the experiment.}\label{Ams_comparison}
\end{table}

\subsection{Conclusions}

In summary, let us formulate the main theoretical results in the hyperfine structure calculations of the HD$^+$ molecular ion that were obtained in recent years:
\begin{itemize}
\item The spin-spin scalar interaction coefficients ($E_4$ and $E_5$) are now available with relative precision of $\sim\!10^{-6}$ \cite{spin-spin20}.
\item The spin-orbit coefficient $E_1$ is obtained with corrections up to $m\alpha^7\ln(\alpha)$ order (this work).
\item The spin-spin tensor coefficients $E_6$ and $E_7$ have been obtained up to $m\alpha^6$ order. Improved numerical results for these coefficients will be presented in a forthcoming publication.
\item Other HFS interactions (not included into the effective HFS Hamiltonian (\ref{HFS_eff_h})) were also studied, particularly the nuclear spin-spin interaction mediated by the electron spin \cite{Ramsey}. It is found \cite{FFK21} that this correction is too small to explain the observed discrepancies.
\end{itemize}
In a view of these achievements, it can be stated that the favored hyperfine components of ro-vibration transition lines can now be obtaied with a theoretical precision that is mainly limited by the calculation of the spin-averaged transition frequency, i.e., $1.4\times10^{-11}$ for pure rotational transitions and $7.5\times10^{-12}$ for vibrational transitions. The main sources of theoretical uncertainty are the $m\alpha^8$ order one- and two-loop contributions to the spin-averaged transition frequency \cite{Korobov17,Korobov21}.

\subsection{Acknowledgements}

This work was done in collaboration with Laurent Hilico, Mohammad Haidar (LKB), and Zhen-Xiang Zhong (Wuhan, WIPM CAS) that is gratefully acknowledged.

\end{document}